\documentclass[12pt]{article}
\usepackage{graphicx}
\usepackage{epsfig}
\usepackage{amsmath}
\usepackage{amssymb}
\usepackage{amsthm}
\usepackage{graphics}

\newcommand{\lp}{\left}
\newcommand{\rp}{\right}

\newcommand{\be}{\begin{equation}}
\newcommand{\ee}{\end{equation}}

\begin{document}

\begin{flushright}
CERN-PH-TH/2008-250
\end{flushright}
\vskip 1cm

\centerline{\Large \bf Micro Black Holes}%

\vspace{.5cm}

\centerline{\Large \bf and the Democratic Transition }%

\vspace{1cm}

\centerline{\large Gia Dvali$^{a,b}$ and Oriol Pujol{\`a}s$^b$ }

\vspace{5mm}

\centerline{\emph{$^a$CERN, Theory Division, CH-1211 Geneva 23,
Switzerland}}

\centerline{\emph{$^b$Center for Cosmology and Particle Physics}}
\centerline{\emph{New York University, New York, NY 10003, USA} }

\vspace{1cm}

\begin{abstract}
Unitarity  implies that the evaporation of microscopic
quasi-classical black holes  cannot be universal in different
particle species. This  creates a puzzle,  since it conflicts with
the  thermal nature of quasi-classical black holes, according to
which all the species should see the same horizon  and be produced
with the same Hawking temperatures. We resolve this puzzle by
showing that for the microscopic black holes, on top the usual
quantum evaporation time, there is a new  time-scale which
characterizes a purely classical process during which the black
hole looses the ability to differentiate among the species, and
becomes democratic. We demonstrate this phenomenon in a
well-understood framework of large extra dimensions, with a number
of parallel branes. An initially non-democratic  black hole is the
one localized on one of the branes, with its high-dimensional
Schwarzschild radius being much shorter than the interbrane
distance.  Such a black hole seemingly cannot evaporate into the
species localized on the other  branes, that are beyond its reach.
We demonstrate that in reality the system evolves classically in
time, in such a way that the black hole accretes  the neighboring
branes. The end result is a completely democratic static
configuration,  in which all the branes  share the same black
hole, and all the species are produced with the same Hawking
temperature. Thus,  just like their macroscopic counterparts, the
microscopic  black holes are universal bridges to the hidden
sector physics.

\end{abstract}

\newpage

\tableofcontents

\newpage

\section{Non-Democracy Puzzle }

By now,  it is understood \cite{bound,cutoff}, that from the
consistency of the large-distance black hole (BH) physics  it
follows, that  in any effective field theory coupled to Einsteinian
gravity, there is the following ultimate connection between the
gravity cutoff
$M_*$ and the number $N$ of the particle species below it, %
\be%
\label{cutoff}
 M_{*}  \, = \, M_P/\sqrt{N} \,.
\ee%
 The physical meaning of  the parameter  $M_*$ is similar to
that of  $M_P$ in ordinary gravity with $N\sim 1$.  It marks the
fundamental scale at which gravity becomes strong. In particular,
it sets the lower bound on the size and the mass of a black hole that can be treated
quasi-classically. For example, any sufficiently compact
source of mass $\gg M_*$ represents a quasi-classical BH with a generalized Schwarzschild radius
$r_g \, \gtrsim \, l_*\, \equiv \, M_*^{-1}$.

Perhaps, the simplest proof of the relation (\ref{cutoff}) comes
from  the consistency of  BH evaporation \cite{bound, cutoff}, but
exactly the same bound follows from a number of other BH-related
fully non-perturbative  arguments, such as the BH entanglement
entropy \cite{ententropy} and the BH quantum information
considerations in the presence of species \cite{quantumN}. Another
indication comes from the perturbative renormalization of the
gravitational coupling \cite{greg,veneziano,cutoff}.  In our
analysis, however,  we shall only focus on the former exact
non-perturbative treatments, which are immune against the
artifacts of the perturbation theory, such as, e.g., cancellations
among the different  contributions.

  We shall now briefly reproduce the BH evaporation argument.
Consider a theory with $N$
light  species $\Phi_j$, where $j\, = \, 1,2,...N$ is the species
label. Let us show that the gravity cutoff in such a theory cannot
exceed (\ref{cutoff}).   We can prove this  by showing  that the
opposite assumption will inevitably lead us to a contradiction.
Thus, let us assume the opposite,  that the cutoff $M_*$ is  much
above the scale $M_P/\sqrt{N}$. Then, at distances  $\gg \, l_*$
gravity, by default, must be in the classical Einsteinian regime. In
particular, a  BH  of an intermediate size,
\begin{equation}
\label{rg} l_*\,  \ll  \, r_g \,  \lesssim \, \sqrt{N}/M_P \, ,
\end{equation}
is also in a quasi-classical regime. The quasi-classicality  implies
that such a  BH should evaporate on the time-scale much longer
than its size $r_g$.  If this condition is not satisfied the rate
of the temperature-change exceeds the
temperature square, which would imply that such a BH cannot be
considered as a thermal state with a well-defined Hawking
temperature $T$, and thus, cannot be quasi-classical. For
quasi-classicality of any given BH, the necessary condition is,
\begin{equation}
{dT \over dt} \, \ll \, T^2 \, . \label{temperature}
\end{equation}
But this condition is impossible to satisfy for any Schwarzschild
BH of size $ \lesssim \sqrt{N}/M_P $. Indeed,  if the black hole
were quasi-classical and Einsteinian,  its half-evaporation time would be
\begin{equation}
\tau_{BH} \, \sim \, r_g^3M_P^2/N\,  \lesssim \, r_g \, ,
\end{equation}
where the last inequality is obtained by taking into the account
(\ref{rg}). Thus, the black hole has a half-lifetime of order or
smaller than its inverse temperature! The quasi-classicality
condition (\ref{temperature}) is inevitably violated,  which is a
clear indication that such a black hole cannot be regarded as a
quasi-classical state, with a well-defined  temperature.  Thus, we
are lead into  the contradiction with our initial assumption,
that the black holes of size $r_g \ll \sqrt{N}/M_P$ are normal
Schwarzschild black holes. The only resolution of this
inconsistency is that the gravity cutoff is (\ref{cutoff}).
Notice, that this conclusion is absolutely insensitive to what
happens to the black hole at the later stage. It is unimportant
whether  the black hole evaporates completely or whether  some new
physics sets in and stabilizes the black hole at some size $\ll
l_*$.\\

A well known example in which the relation  (\ref{cutoff}) is
manifest from the fundamental physics, is the framework of
large extra dimensions.  If we consider a  $4+n$-dimensional
theory with  a fundamental Planck mass  $M_*$, and  $n$ extra
dimensions compactified  on a certain manifold, the relation
(\ref{cutoff}) simply translates as the familiar  relation between
the $n+4$-dimensional  and $4$-dimensional Planck
masses,
\begin{equation} \label{planckscales1} M_P^2 \, = \, M_*^2
\, V_{(n)} \, ,
\end{equation}
where $V_{(n)} \, = \, N_{KK}$ is the volume of the extra space
measured in $l_*$ units. Obviously, this quantity  counts the
effective number of four-dimensional Kaluza-Klein (KK) species
with masses $\lesssim M_*$.

For example, consider a $4+n$-dimensional theory with  $n$ space
dimensions compactified on an $n$-torus, and $4$ non-compact
dimensions forming the Minkowskian geometry.  Without affecting
any of our results,  for simplicity, we shall set all the compactification radii
being equal to $R \, \gg\, l_*$. The fundamental high-dimensional
theory  has one dimensionful  parameter, the $4+n$-dimensional
Planck mass $M_*$,  which is the cutoff of the theory.  Assume
that the only species in the theory is a $4+n$-dimensional
graviton. From the four-dimensional point of view the same theory
is a theory of the tower of spin-2 KK species. For any $R$,  the
relation between the four-dimensional Planck mass and the cutoff
of the theory is \cite{add}
\begin{equation}
\label{planckscales} M_P^2 \, = \, M_*^2 \, (RM_*)^n \, .
\end{equation}
 The key fact establishing the connection between (\ref{planckscales}) and (\ref{cutoff}) is
that the factor $(RM_*)^n$ measures the number of KK species,
\begin{equation}
\label{kknumber} N_{KK} \, = \, (RM_*)^n \,.
\end{equation}
Thus,  the relation (\ref{planckscales}),  is a particular example
of relation (\ref{cutoff}) in which $N$ has to be understood as
the number of KK-species.\\

As pointed out in \cite{GiaBHs}, the fact that in $N$-species
theory the gravity cutoff is much below $M_P$, inevitably implies
the existence of a second length scale, $\geqslant l_*$,  which
marks the distance at which  gravity starts departing from the
Eisteinian regime. We shall denote this scale by $\mathcal{R}$.

In the interval  $l_*  \, \ll \, r \, \ll \, \mathcal{R}$,
gravity is still in a weakly-coupled classical regime, but it is
{\it non-Einsteinian}.  We define non-Einsteinianity in the
obvious sense, that at such distances gravitational interaction is
no longer mediated by a single massless spin-2 state, but also by
some new degrees of freedom. As a result, in general, neither the
gravitational potentials nor the BH parameters are obliged to obey
the usual Einsteinian or Newtonian laws.

Correspondingly, the black holes with $r_g$ in this interval are
still classical,   but non-Einsteinian. In large extra dimensional
example the length-scale $\mathcal{R}$ has an obvious meaning, it
is  the same as the radius of extra dimensions $\mathcal{R} \,
\equiv \, R$, beyond which the classical gravity becomes
high-dimensional. Obviously, the BHs of the intermediate size $l_*
\, \ll \, r \, \ll \, \mathcal{R}$ are still classical,  but
high-dimensional.  In particular the gravitational radius of such
a BH is related to its mass ($\mathcal{M}$) in the following way,
\begin{equation}
\label{nrg} r_g \, \simeq\, l_* (\mathcal{M}\,l_*)^{{1 \over n+1}}
\, .
\end{equation}

\begin{figure}[t]
\vspace{-2cm}
\begin{center}
  \includegraphics[height=8cm]{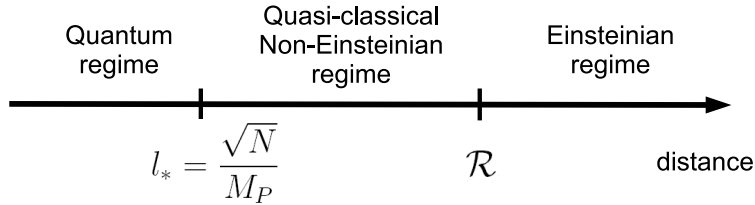}%
\end{center}
\vspace{-3cm} \caption{Regimes of gravity in the presence of $N$
species. For the particular case of large extra dimensions ($N =
N_{KK}$), the gravitational cutoff is the higher dimensional
Planck length $l_*=M_*^{-1}$ and the scale at which the first
deviations from Einteinian gravity show up is the compactification
radius $R$.}
\label{fig:chunk} %
\end{figure}

Notice that in certain $N$-species theories the two scales
$\mathcal{R}$ and $l_*$ can be close to each other, but this
proximity is an illusion.  To understand this, it is useful to go
into the energy-momentum space instead.  Even if the length-scales
are close to each other, the mass interval in which the  BHs are
in non-Einsteinian regime is nevertheless huge,
\begin{equation}
\label{interval} M_* \, \ll \, \mathcal{M}_{{\rm non-Einstein}}
\, \ll  \, \mathcal{R}M_{P}^2 \, .
\end{equation}
The lower bound of this interval is obvious, and the upper bound
comes from the fact that for the Einsteinian  BHs of gravitational
radius $r_g \, = \, \mathcal{R}$ the mass and the size are related
through the usual  Schwarzschild  formula. As an illustrative
example, consider a Kaluza-Klein theory with $n \gg 1$ extra
dimensions. Obviously, in the large $n$ limit compactification
radius approaches $l_*$ from above. Nevertheless, the volume of
extra space stays constant in units of $l_*$. A high-dimensional
BH, becomes Einsteinian  after its $4+n$-dimensional gravitational
radius (\ref{nrg}) becomes equal to the  $4$-dimensional
gravitational radius of a usual Schwarzschild BH with the same
mass. This happens when $r_g \, = \, R$, that is,  only after the
BH  fills up the whole high-dimensional volume, and thus it has to
be
very massive.\\

The existence of the two critical length scales $l_*$ and
$\mathcal{R}$ in the light of unitarity and large-distance BH
properties, leads us to the two powerful conclusions:

 $~~~~~~~$

{\bf  (1)} The evaporation of the quasi-classical microscopic BHs
with the size $\ll \mathcal{R}$, is non-democratic in species.
That is,  the micro BHs carry `hair' that distinguishes among the
different species.

$~~~~~~~$

{\bf (2)}  The democracy is gradually  regained  for the larger
BHs, and becomes complete for  $r_g \, > \, \mathcal{R}$.

 $~~~~~~~$

In order to understand the above properties, it is enough to
consider a thought experiment with BH production and subsequent
evaporation. A neutral BH of mass $\mathcal{M} \, \geqslant  \,
M_*$ and gravitational radius $r_g$, can be produced in a
particle-antiparticle collision within  the given (say, $j$-th)
species, at the center of mass energy $E \sim \mathcal{M}$ and an
impact parameter $R_{imp} \sim r_g$. At the threshold of the
smallest BHs creation,  the dynamics is governed by a single mass
scale   $E \sim R_{imp} \sim M_*$, and the production rate
obviously is  $\Gamma_{j \rightarrow BH} \, \sim \, M_*$. By
unitarity, the rate of BH  decay  into a pair of the same $j$-th
species should be similar
\begin{equation}
\label{ratej} \Gamma_{BH \rightarrow j} \, \sim  \,  M_* \, .
\end{equation}
But then,  by the same unitarity,  it is impossible for the BH to
have the same partial decay rates into the other species
individually. Since the total rate of decay into all the other
species cannot exceed $\sim M_*$,
\begin{equation}
\label{ratetotal} \sum_{i\neq j} \Gamma_{BH \rightarrow i} \,
\lesssim  \, M_*,
\end{equation}
the decay rate into most of the other individual species must be
suppressed by $\sim 1/N$ factor. Thus, the decay of the
microscopic BH {\it cannot be democratic}, and the original BH has to
carry some information that allows to distinguish among the
different species.  Since the BH by construction was neutral, this
information cannot correspond to any conserved charge measurable
at infinity.   In the other words, we are lead to the conclusion
that the micro BHs must carry \emph{species hair}.

At the same time, the level of  non-democracy is size-dependent.
By increasing the gravitational radius gradually, we must recover
the complete democracy at $r_g \, = \, \mathcal{R}$, since above
this size the BHs become Einsteinian.

Both properties stated above acquire a clear geometric meaning in
case of the extra dimensional example.  Indeed, imagine an extra
dimensional theory with $n$ dimensions compactified  on a manifold
of radius $R \, \gg \, l_*$, and with two parallel $3$-branes
placed at some intermediate distance $d$ ($l_* \, \ll \, d \, \ll
\, R$) from each other.  Both branes are point-like on the compact
$n$-dimensional manifold. Assume that there are two different
$4$-dimensional particle species $\Phi_1$ and $\Phi_2$ localized
on the first and the second branes respectively.

Now, we can form a quasi-classical black hole in a collision (or a
collapse) of species $\Phi_1$ on the first brane.   If the size of
the BH is $r_g \, \ll \,  d$, there is no way for this BH to
evaporate into the species $\Phi_2$ that are localized on the
second brane.   Seemingly, this fact is pretty natural from the
point of view of the five-dimensional observer, and is entirely
due to locality in the extra dimensional space.  Yet,  from the
point of view of a four-dimensional observer the same effect is
rather puzzling. For such an observer, measuring the
four-dimensional  Hawking flux at large distances, the flux will
include only $\Phi_1$ species and this fact would be hard to
reconcile with the thermal nature of the quasi-classical BHs.

This puzzle is not limited to extra dimensional example only, and
is generic for BHs in theories with $N$-species. The non-democracy
of the BH evaporation implies that effectively such BHs have
different temperature with respect to the different species, but
this contradicts the notion of thermality.

The ability of a BH to distinguish among the different species is
also puzzling from the point of view of the BH no-hair theorems
\cite{nohair}. Indeed, a non-democratic BH can be labeled
according to its interaction with species, and thus, it carries a
{\it species hair}.   However, since in the above thought
experiment the BH produced in the collision of particles was
neutral {\it by design},  its species hair cannot correspond to
any conserved quantum number that can be detected from outside.
This fact goes in contrast with the no-hair properties of the
Einsteinian BHs \cite{nohair}, which can only be labeled by
charges that can be detected at infinity, either classically or
quantum-mechanically \cite{quantum}.

A quick and naive resolution (or rather a dismissal) of the above
puzzle would be to argue, that, because the BHs of the size $r_g
\, \ll \, \mathcal{R}$ are non-Einsteinian, they have no
obligation to obey the usual universal thermal rules.
Nor they should be subjected to  the classical no-hair theorems.
 However,
such reasoning is incorrect, since the thermal nature of BHs is a
result of {\it quasi-classicality} rather than of {\it
Einsteinanity}. The former requirement is obviously satisfied in
the region of the interest. For example, the high-dimensional BHs
that are smaller than the compactification radius, but larger than
the fundamental Planck length, are obviously non-Einsteinian, but
they are still quasi-classical. Such BHs must be the thermal
states. Then, how can such BHs be non-democratic in species?

 It is the purpose of the present work to answer this question. We
shall argue that the resolution of the puzzle is in the
time-dependence of the non-democratic BHs. In the other words, the
microscopic BHs can exhibit non-democracy only for a limited time
interval, after which they lose the ability to distinguish among
the species and become fully democratic. This democratization is a
classical process.  In the other words, the species hair is
time-dependent and falls off within a finite time.  At the end of
this process,  all the interactions between the BH and the species
are universal, and there is a single temperature with respect to
all of them. Thus, we conclude that non-democracy is only a
property of classically-time-dependent micro BHs, and the static ones are
fully democratic. This fact reconciles the non-democracy with the
known properties of quasi-classical BHs.  Both the absence of hair
as well as fully thermal nature are only the exact properties of the static BHs.

In the present paper we shall demonstrate all the above properties
on the explicit extra dimensional example with separated branes
discussed above.  The classical configuration in which a BH is
localized on a single brane cannot be static. What happens is that
the system evolves classically until all the branes `share' the
same BH, after which all the modes see the same horizon, and are
produced with the  same Hawking temperature. In the other words,
brane \emph{accretion} by the BH takes place.

In the usual extra dimensional setup there are many factors (such
as stabilization of branes and extra dimensions, resolution of the
brane thickness etc)  which one usually fixes by some simplifying
assumptions,  without worrying about the internal consistency of
the underlying  setup.  For some effective field theory
treatments, such assumptions are fully justified, but the effect
we are considering may be very sensitive to the  intrinsic
consistency of such constructions. So we perform our study in the
setups which we can fully control.

We consider various examples of large compact extra dimensions in
which some four-dimensional species are localized on the 3-branes.
We then ask the question, whether it is possible to have a stable
classical BH, that will {\it not} evaporate at least in some of
the species. We attempt to create such a situation by localizing a
classical BH away from some of the branes, and in this way
isolating it from the species that populate the latter. The naive
expectation then is, that the BH in question will not be able to
evaporate into the distant species, simply by the locality in the
extra space. However, we find that this never happens.  The
consistent gravitational dynamics is always such that the BH finds
the way to get in contact with all the existing species within the
finite time!  The end result, is always a fully democratic BH.

For example, in co-dimension one case, we localize a classical BH
away from the brane by attaching it to the latter by a string
(flux tube), using the setup of \cite{giasergey}. By choosing the
appropriate string tension one can balance the gravitational
repulsion created by the brane with the attractive force due to
stretched string and stabilize the BH in a neutral equilibrium at
an arbitrarily large distance. In this way one can seemingly
isolate the BH from the modes localized on the brane. However,
this is not what happens. We show, that in reality, it is
energetically favorable for the string to shorten and be replaced
by a chunk of the brane itself (see Figure \ref{fig:chunk}). As a
result, the BH is attached to the brane by the deformed brane
portion, and all the brane modes are produced democratically.

\begin{figure}[t]
\vspace{-0.5cm}
\begin{center}
  \includegraphics[height=6cm]{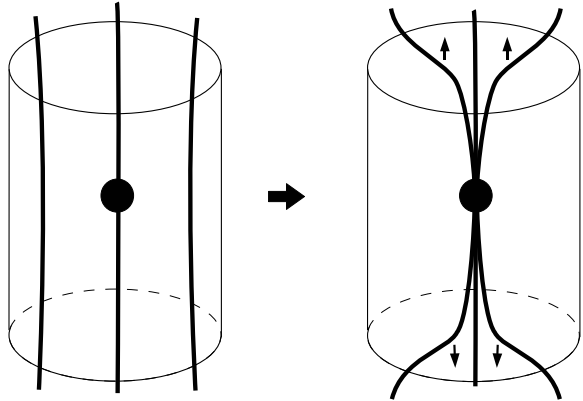}%
\end{center}
\vspace{-5mm} \caption{Once a Micro BH is formed in one of the
branes, it attracts and accretes the neighboring branes with a
time-scale that depends on the initial interbrane separation. The
end result is that all the branes intersect the BH and so the
evaporation proceeds democratically, and so the species hair is
lost. This is accompanied by a wave of the brane fluctuation modes
that expands to infinity in the 4D directions, and corresponds to
the species hair being expelled from the BH.}
\label{fig:accretion} %
\end{figure}

The second example, is the setup with parallel codimension-2
3-branes in six dimensional space time. The transverse metric
produced by these objects is very similar to the one produced by
the cosmic strings in  four dimensions, and amounts to a
locally-flat space with a conical deficit. As long as the deficit
angle is less than $2\pi$, the four-dimensional metric produced by
these defects is Minkowskian. We consider a BH that initially is
localized on one of such branes, so that only the modes from the
piercing brane see the BH horizon and can be emitted thermally. We
show that the end result of the classical evolution of the system
is that the BH is pierced through by all the existing branes
symmetrically, and thus, all the localized modes
see a common BH horizon and temperature. This process is
described in Figure \ref{fig:accretion}.\\

The results of this paper have phenomenological implications for
the microscopic BHs that can be created at LHC, since they
indicate that such BH will have an intrinsic characteristic
classical time scale on which they lose the species hair and
become democratic. Although non-democracy turns out to be a
temporary property,  for the microscopic BHs it plays the crucial
role in maintaining unitarity.
This is because, in the presence of many species, for the smallest micro BHs the classical time of full
democratization is usually much longer than the quantum
evaporation time.  Such BHs  for all the practical purposes will appear
to be non-democratic.  However,  with the increasing mass of the BH the evaporation
rate decreases whereas the democratization process becomes more efficient.
The interplay between the two processes is the characteristic property of any micro BH, and
this fact has important phenomenological implications for the microscopic BHs that can be
produced at LHC.

 The  democratization phenomenon, discussed in this paper,  demonstrates  that the universality is
 not exclusively a property of the Einsteinian classical BHs, but is  shared also by the
 microscopic non-Einsteinian ones.  Thus, just like their macroscopic counterparts, the micro BHs  represent  universal bridges to the hidden sector physics.

\section{The Principle of Micro Black Hole Democracy}

We shall now formulate the principle of micro-BH democracy, explicit
manifestations of which will be discussed in the following sections.

We shall restrict our considerations to backgrounds that at large
distances represent asymptotically-flat four-dimensional Minkowski
spaces.  On such a background, let us consider an arbitrary
classically-static and stable BH configuration,  for which the
back-reaction from the Hawking radiation is small (i.~e.,  such a
BH can be consistently treated as a  quasi-classical state). Then,
all the species that can be treated as the point-like elementary
particles, at least, up to the distances of order the BH size, are
produced in the evaporation of the BH at the same temperature.
What makes the above formulation more general with respect to the
usual universality of the Hawking radiation for macroscopic BHs,
is the absence of the assumption that gravity is Einsteinian.
Instead, what is important, is that the BHs are the
quasi-classical states, implying that they correspond to stable
static solutions of a classical gravity theory operating at the
intermediate distances $l_* \, \ll \, r_g \, \ll \, \mathcal{R}$,
for which the quantum back-reaction is small. This theory by
default is not Einsteinian gravity, but rather its classical
short-distance completion, such as, for example, a large extra
dimensional theory  of gravity.

Applying to such completions,  the above principle will suggest,
that any four-dimensional zero mode, with the localization width
shorter than the high-dimensional BH gravitational radius,  must
be produced with the same Hawking temperature.   In other words,
any static quasi-classical BH must be seen by all the zero modes
in the same way.

This is a pretty strong statement, which naively seems to be
easily circumvented due to locality in the extra space. Indeed, it
would  imply that no static configuration, in which modes are
localized away from the position of a high-dimensional BH, is
possible.  Nevertheless, things are more subtle, as we shall see.

We shall now consider some seeming counterexamples, in which, at
least naively, the locality in high-dimensions should prevent the
localized modes from being produced in the BH evaporation,  and
show why what happens is exactly the opposite.

\section{Co-Dimension One Case}

It has been known for a long time \cite{vilenkinCS,vilenkin}, that
gravity of positive tension codimension-one branes (domain walls)
is repulsive. So naively, it seems straightforward to stabilize a
classical BH away from the brane at an arbitrarily large distance,
by balancing the repulsive gravity by some attractive force. One
would expect then, that in such a case the BH would not be able to
evaporate in the modes that are localized on the brane, in
contradiction to our democracy principle. We shall now discuss why
this naive intuition is false.

 The relevant mechanism for stabilizing a classical BH away from the brane
was discussed in \cite{giasergey}.  So let us  follow this
construction.  We shall discuss the case of a positive
tension $3$-brane, with the energy momentum tensor $T_{A}^{B} \, =
\, T \; {\rm diag} (1,1,1,1,0)\, ~~~ T \, > \, 0$, embedded in
five-dimensional space-time.  The solution with a
locally-Minkowskian bulk metric can be written as\cite{vilenkin}
\begin{equation}
ds^2 \, = \, (1 -|y|\kappa)^2 [dt^2 \, - \, e^{2 t\kappa}dx^2]\, -
\, dy^2 \,, \label{wallmetric}
\end{equation}
where $\kappa \, = \, T / (3\, M_*^3)$ is the gravitational
curvature radius of the brane. Due to the decreasing warp-factor,
the force exerted by the brane on test particles is repulsive in
the $y$-direction. However, the four-dimensional induced metric is
not Minkowski but rather de Sitter with the Hubble radius set by
$\kappa$. In order to create a static metric, one needs a source
that would absorb the gravity flux, and such a source is the
negative  bulk cosmological constant $\Lambda$. By carefully
tuning $\Lambda$ versus the brane tension, in such a way that
$\kappa$ coincides with the bulk AdS curvature scale
\begin{equation}
\kappa \, = \sqrt{|\Lambda| \over 6M_*^3} \label{tuning}
\end{equation}
 one can create a configuration, with a static
4D Minkowski metric \cite{RS2}
\begin{equation}
ds^2 \, = \, e^{-2 |y|\kappa} (dt^2 \, - \,dx^2)\, - \, dy^2 \,.
\label{RS}
\end{equation}
In such a metric, a BH is repelled from the brane and it looks
like we can create a counterexample, provided we stabilize the BH
at some finite distance. We shall now discuss the method of
stabilization considered in \cite{giasergey}.

\subsection{Step One: Compactification}

In order to obey the conditions of our micro-BH democracy
principle, we first need to obtain the four-dimensional gravity at
large distances. We thus need to compactify the extra space, that
is that the  $y$ coordinate in \eqref{RS} has a finite range.
However, as it is known \cite{RS1}, such compactification requires
the introduction of a negative tension brane, say,  at $y=y_0$.
The latter requirement on the tension directly follows from the
matching of the warp-factor, at the brane location. In
\cite{giasergey} it was noticed that by increasing $y_0$, such a
setup allows to create a quasi-classical BH even for a small BH
mass, due to blue shift of the local Planck length $l_{local} \, =
\, l_* e^{|y_0|\kappa}$. In order to be in a five-dimensional
quasi-classical regime, the BH size must be smaller  than the 5D
AdS curvature radius ($\kappa^{-1}$),  but larger than the local
(blue-shifted) Planck length,
\begin{equation}
\kappa^{-1} \, \gg \, r_g \, \gg \, l_{local} \, .
\label{condition}
\end{equation}

In such a case, it seems that the BH can be easily localized far enough
from $y=0$ and cannot emit radiation in the
modes localized at the positive tension brane, in seeming contradiction
with our claims.
Before declaring victory however, one has to address a crucial subtlety, which
as we shall see, will reverse the above naive conclusion.

 The negative tension brane can only be
considered as an effective description at large distances. But in
order to fully understand its possible effect on the nearby
localized BHs, this large distance description does not suffice.
The negative tension brane has to be resolved by some short
distance physics,  and the BH behavior in its vicinity will
crucially depend on the nature of this regulating physics.
Currently, the dynamics of such resolving physics is not well
understood. Therefore, to be on the safe side in our analysis, we
will stabilize the BH far away from the negative tension brane, in
the region that is not influenced by the ultra-violet physics that
resolves the latter.

\subsection{Stabilization by the Stretched String}

\begin{figure}[t]
\begin{center}
  \includegraphics[height=8cm]{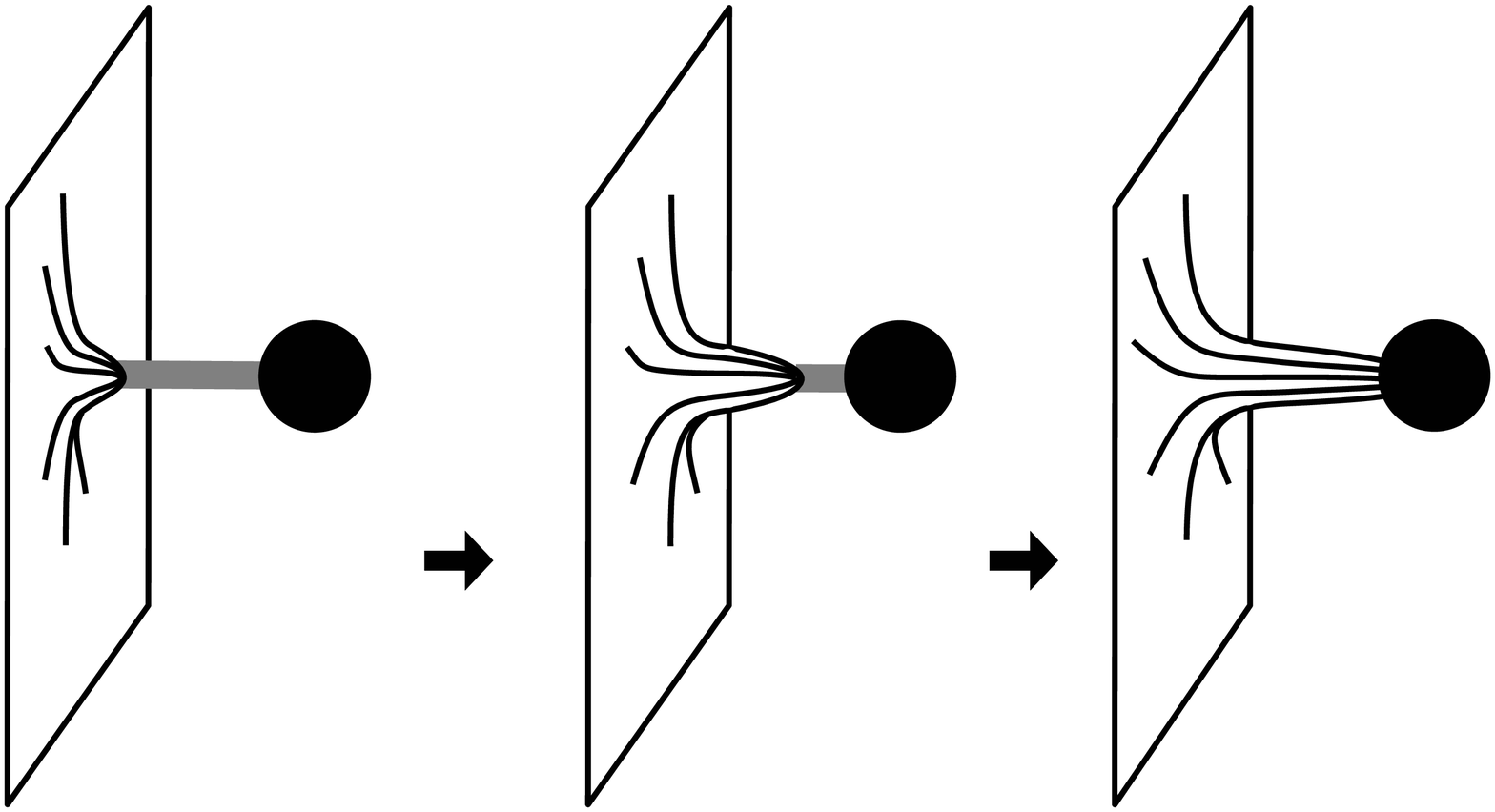}%
\end{center}
\vspace{-2cm} \caption{~}
\label{fig:chunk} %
\end{figure}

Ref. \cite{giasergey} also suggested an alternative way of
stabilizing the BH away from the positive-tension brane, by
attaching the BH to the latter by a string. The string in question
can  represent either a QCD-type electric flux tube, or be
solitonic. It could even be a fundamental string.  The precise
short-distance nature is unimportant for our purposes. The
explicit realizations of the above setup were discussed in
\cite{giasergey}.  For example,  appearance of such a string is
automatic whenever the BH carries a charge under a massless  gauge
field localized on the positive tension brane \cite{dvalishifman},
and it is an inevitable consequence  of charge conservation in 4D.
In the latter case, the string in question is an electric flux
tube. For our present purpose, this method of the BH stabilization
has a clear advantage, that we can avoid any encounter with the
negative tension brane physics.  Choosing a small $\kappa$, one
can place a negative tension brane  arbitrarily far, while keeping
the BH attached to the positive tension brane at a fixed distance.
To obtain  the static configuration we have  to balance the
repulsive gravitational force acting on the BH, by the attractive
one induced by the tension of the stretched string. This amounts
to the following condition
\begin{equation}
\mu \, = \, \kappa\, M_{BH}\, , \label{balance}
\end{equation}
where,  $\mu$ is  the string tension,  and $M_{BH}$ is the BH
mass. Under the above condition, the BH is in a neutral
equilibrium and can be placed at an arbitrary distance away from
the brane.  Such a configuration, however, is exactly  in
agreement with the BH democracy, since the latter will evaporate
through the modes localized on the brane, because the Hawking
radiation from the BH is conducted to the brane through  the
string.  As a result,  a 4D brane observer will see  the Hawking
radiation in form of the brane modes. Notice,  that  the condition
(\ref{balance}) can be rewritten in the following way,
\begin{equation}
\mu^2 \, = \, (r_g^2\mu) \,  T \, . \label{balance1}
\end{equation}
Taking into the account that the string and brane widths are
$\varepsilon_{brane} \, \sim \, T^{-{1\over 4}}$ and
$\varepsilon_{string} \, \sim \, \mu^{-{1\over 2}}$, we see that
when the BH is wider than the string, the string tension exceeds
the brane tension. So it is energetically favorable  for the
string to shorten and pull out the  stringy chunk of  brane (see
Fig. \ref{fig:chunk}). As a result,  the original string, will be
replaced by the stretched part of the brane, and the BH  becomes
attached to the brane by the tube of the stretched brane itself.
Hence, all the brane modes are produced democratically.

In the opposite case, when $r_g^2 > \mu$,  both brane and string
are wider than the BH, and the conditions of our principle  are
automatically  violated, since the mode localization-width becomes
wider than the BH size.

\section{Co-Dimension Two Case}

The second piece of evidence in favor of the Principle of Micro BH
Democracy arises from a model with two compact extra dimensions
and a number of codimension-2 branes placed at different locations
in the extra space. From the 4D perspective, the modes localized
in each of the branes are species that only interact though
gravity. Naively, it seems that it should be possible to have a stable situation in which
a mini BH is pierced by only one of the branes. Then, by the locality in the
extra space, only the modes localized on that brane would be thermally
produced by the BH, which from the 4D point of view would look
like a BH evaporating into only some of the species. Hence,  such a BH would
carry a classical  {\it species hair}.

We shall now see that the classical dynamics drives the branes +
BH system to a configuration in which the BH becomes  pierced by all the
branes.  This transition happens within  a finite time. As a result, the modes localized on all
the branes see the same horizon and are emitted with the same temperature. Thus,  the species
hair is lost classically, in a characteristic time that depends on
the model.

We demonstrate this by finding the attractive force between the BH
and the codimension-2 3-branes.  Because the  gravitational
interaction of codimension-2 branes is very similar to that of
cosmic strings  in 4D,  from now on we shall simply refer to them
as `Cosmic Strings' (CSs). It is well known \cite{vilenkin} that a
CS produces no (static) force on test particles. However, this is
not the end of the story, since there are nontrivial higher order
effects that become important both for a BH and for a point
particle, whenever one goes beyond the `test particle' limit.

\subsection{Brane-BH interactions}

Let us now find the force between a BH and a codimension-2 brane.
It is straightforward to see that the force exerted by the Cosmic
String (CS) on a test particle  at linear level in $G_N$
 vanishes. This follows from the single graviton exchange
amplitude, which in $D$ dimensions is
$$
{\cal A} \sim \;G_N\; \int d^Dx\; {T'}^{MN}\, {1\over \Box_{D}}
\,\lp(T_{MN}-{1\over D-2}\,T^R_R\;\eta_{MN} \rp) ~,
$$
where $T_{MN}$ and $T'_{MN}$ are the stress tensors of the
codimension-2 brane and of the point particle (or vice-versa) respectively.
The above amplitude is automatically zero for a codimension-2 brane, because
$T_M^N$ has $D-2$ nonzero and equal entries. This fact  can also be seen
from  exact solution representing a straight CS, which is flat
space with a wedge removed \cite{vilenkinCS}. This solution is
locally flat and has a zero Newtonian potential.

However, there is a non-zero attractive force at order $G_N^2$. In
fact, it is possible to identify two independent contributions to
the force at this order. The first was discussed long ago in the
context of Cosmic Strings in 4D
\cite{linet,smith,galtsov,vachaspati}, and even though it is a
$G_N^2$ effect it does not account for the graviton
self-interactions. Rather, this effect arises because the CS
imposes non-trivial boundary conditions.
%These are such that a point
%particle feels the presence of a number of image particles
%distributed around the CS, and as a result a force towards the CS
%appears.
We describe this phenomenon in more detail in Section
\ref{sec:CSbackground}.

The other $O(G_N^2)$ contribution to the CS-BH force arises by
taking into account the spacetime curvature induced by the BH (or
by the localized particle, the exterior solution of which is a BH
metric). This can be most easily extracted by considering a probe
CS on a fixed BH background, as was done in \cite{escape2} (see
also \cite{escape1}). For completeness, we review this computation
in Section \ref{sec:BHbackground} (see also Appendix
\ref{sec:braneBHpot}). Let us only add now that one can view this
contribution to the force as arising from a two-graviton exchange
diagram with one graviton vertex. Hence, this contribution arises
from the nonlinear structure of higher-dimensional GR. In
contrast, the previous contribution has nothing to do with the
nonlinearities of GR, as it would be present even in a linear
gravity theory.

Finally, in Section \ref{sec:full} we further argue that the
end-result of the BH-CS system must be the configuration in which the BH `eats up' a
segment of the CS. This result is supported by the thermodynamical properties of the
known exact solutions representing a BH pierced by a CS
\cite{afv}.

\subsubsection{Interacting with the images}
%\addcontentsline{toc}{subsubsection}{Interacting with the images}
\label{sec:CSbackground}

As mentioned above, CSs do not exert any force on point particles
in the test limit,  in  which any gravitational effect of the
particle is neglected.  A simple way to go beyond this limit is to
compute the gravitational field of the point particle on the exact
CS background  in a quasi-Newtonian approximation, that is,  to
compute the Newtonian potential due to a point source on a conical
space.
This simple logic already leads to an $O(G_N^2)$ attractive force
between a point-like source and a CS
\cite{linet,smith,galtsov,vachaspati} (even though as we argue in
\ref{sec:BHbackground}, this does not capture all the
contributions to the force in the given order).

The key point is that the Newtonian potential $\phi_N$ due to a
point source of mass $m$ in a conical space ({\em i.e.} with a
conical singularity along  the $x$ axis) is equivalent to that
produced by the same source plus a few image sources of the same
mass, with the number of images depending on the CS tension $T$.
This is a consequence of the Gauss' law: in the conical space with
deficit angle $\delta$, the field lines from a particle  spread
through a volume that is `smaller' by a factor $(1-{\delta/
2\pi})$ as compared to without the CS. Hence, the gravitational
field is stronger than for the particle in isolation. The
additional contribution is of course proportional to $\delta$ for
small tensions, and can be thought of as due to a number of images
distributed around the CS.

\begin{figure}[t]
\begin{center}
  \includegraphics[height=7.5cm]{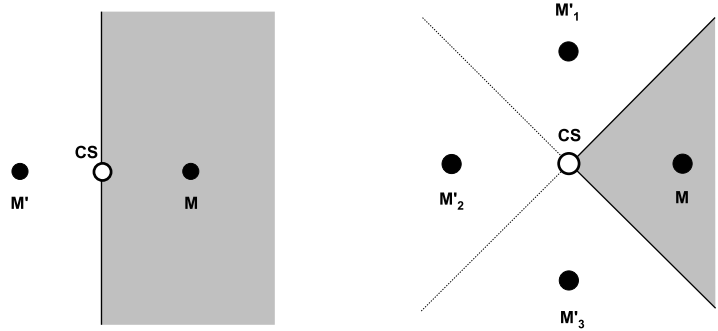}%
\end{center}
\vspace{-1.5cm} %
\caption{The method of images for codimension 2 branes with
deficit angles equal to $\pi$ (left) and $3\pi/2$ (right). The
Green's function with source point labelled by $M$ in the conical
space (shaded area) coincides with that in the whole plane with
images at points $M'_i$ symmetrically distributed around the
Cosmic String.}
\label{fig:images} %
\end{figure}

This is especially clear for the values of the CS tension for which the
deficit angle is of the form
$$
\delta=2\pi{k\over k+1} \qquad\quad k=1, 2, \dots
$$
The resulting conical spaces for $k=1$ and $3$ are represented in
Fig \ref{fig:images}. It is clear that $k$ is the number of
images. For general values of the tension, the image sources are
not so neatly distributed, but still a qualitatively-similar
picture should hold, with an effective number of images given by
$$
{T \over 2\pi M_*^4 - T}~,
$$
where $M_*$ is the 6D Planck mass.

Now, if $\phi_N$ takes the same form of the usual potential plus
the one due to the images, then for all the practical purposes the
picture should be as if the images are there, and in particular,
the point source itself should feel the gravitational potential
from them. For small tensions, this immediately leads to an
attractive force towards the location of the CS
\begin{equation}\label{forceImages}
F \sim {1\over M_*^8}  {T\,m^2 \over z^4}~,
\end{equation}
where $z$ is the perpendicular distance between the CS and the
particle.

This force is analogous to the one that an electric charge $Q$
feels in the presence of a conducting plate. The boundary
conditions imposed by the plate are such as if there were an image
charge on the opposite side of the conductor. Hence, the particle
feels a force exerted from the plate, even if the latter were
electrically neutral in the absence of the particle. This can also
be understood as the fact that the charge polarizes the medium
that makes up the plate so that the external charge interacts with
the induced charge distribution on the plate. Since the induced
charge itself is proportional to the original charge $Q$, the
resulting force is quadratic in $Q$. In our CS-massive particle
example, one could say that the mass induces a gravitational
dipole distribution along the CS, and this is what attracts the
particle towards it.

\subsubsection{Probe Cosmic Strings on BH backgrounds}
\label{sec:BHbackground}

We can obtain another contribution to the CS-BH force by taking the
limit that is opposite  to the one considered in the previous section.
We shall take a probe brane in an exact BH background.
This is a  well defined problem and it allows to extract the different
contribution to the force.

Let us briefly sketch how this computation is done and quote the
result, while leaving the details for Appendix
\ref{sec:braneBHpot}. This method closely  follows \cite{escape2}.
The idea is to compute the energy of the brane configurations
where the brane asymptotes to a fixed perpendicular distance $z$
to the BH. From this, one can extract the potential energy $V(z)$
that the brane stores for every such configuration. Of course, the
total energy stored in each configuration is divergent because the
brane has infinite extent, but the energy difference between any
two configurations is finite.

For a generic $p-$brane in $4+n$ dimensions one finds that the
potential is already nonzero at linear level and behaves as
\begin{equation}\label{VZ}
V(z) \sim - {T \,m \over M_*^{n+2}} {1\over z^{q-2}}
\end{equation}
where
$$
q\equiv n+3-p~
$$
is  the \emph{codimension} of the brane. This leads to a brane-BH
force that is attractive for $q>2$, zero for $q=2$ and repulsive
for $q=1$, as expected.

For codimension 2, the leading non-zero contribution arises at the
next order, and the result is
\begin{equation}\label{Vcod2}
V(z) \sim -{T m^2 \over M_*^{2(n+2)}} {1\over z^{(n+1)}}~.
\end{equation}
This also leads to an attractive force, which is of the same order
as the one due to the images \eqref{forceImages}.

\subsubsection{Full nonlinear problem and thermodynamics}
\label{sec:full}

In general, obtaining the form of the localized BH metric
including the brane gravity is a difficult task. However, the
codimension 2 case is much more tractable and in fact is the only
known example that allows the explicit construction of a localized
brane BH \cite{KaloperKiley}. Indeed, this case is equivalent to a
BH pierced by a Cosmic String, which can be simply obtained by
reducing the range of an appropriate angular coordinate. Focussing
on our 6D example (and the non-rotating case), this leads to a
higher dimensional version of the Aryal-Ford-Vilenkin metric
\cite{afv}
\begin{eqnarray}\label{afv}
ds^2&=&-f\,dt^2+{dr^2\over f} +r^2 \lp[d\theta_1^2 +
\sin^2\theta_1 d\theta_2^2+...+ b^2 \sin^2\theta_1 \sin^2\theta_2
\sin^2\theta_3 \,d\phi^2 \rp]
\cr%
f(r)&=&1-{1\over b} {r_{g}^3\over r^3}\; , \qquad\qquad
b=1-{\delta\over 2\pi}
\end{eqnarray}
where $\delta=8\pi G_N^{(6)}\,T$ is the deficit angle and $r_g$
depends on the BH mass in the usual 6D fashion without the CS,
$$
r_g^3 = {3\over 2\pi^2} G_{N}^{(6)} m ~.
$$
Notice that the Newtonian potential of the solution $(f(r)-1)$
includes the appropriate enhancement from the CS tension, which
can be understood in terms of the images produced by the string,
as discussed in \ref{sec:CSbackground}. Thus, even if the BH mass
itself does not increase when it `swallows' a segment of the CS,
nevertheless the BH produces a stronger gravitational field. For
$\delta\ll1$, this is equivalent  to an increase in mass
$%
% \Delta m \sim
%{8\pi G_{N}^{(6)} T \over 2\pi} \; { 2\pi r_g^3 \over 3 G_{N}^{(6)}}
\sim(\delta/2\pi)\,m= (8\pi/3) \, T\, r_g^3 \,  $,
which is twice the mass of a three-dimensional ball of radius
$r_g$ and energy density $T$, that is in the segment of CS that
the BH naively `swallows'.

Hence, upon accreting a string of tension $T$ the BH radius
increases, by a factor $b^{-1/3}$. Accordingly, the area of the
pierced BH is
$$
S_{cs} = b \; b^{-4/3} S_{isolated} \, ,
$$
where the first factor comes from the smaller range of one of the
angles, and the other `four' to the larger radius.

As a result, in the accretion process the area and the entropy
increases (at least) by a factor $b^{-1/3}$. This is another
indication that the interaction between the CS and the BH is
attractive. In an energy conserving process, the entropy will grow
if the BH accretes the CS, but not otherwise.

%Initially, CS far away from BH with rest energy $E_i$ and
%gravitational mass $M_i=E_i$.
%
%After coalescence, CS on top of the BH with energy $E_f$ and grav
%mass $M_f$.
%
%Let us assume coalescence is adiabatic, ie, $\Delta S=0$ (for
%large enough BH, the entropy and energy that goes into GWs is
%negligible). Then,
%$$
%r_i^{4}=b r_f^4 \qquad E_i^{4/3}= b^{4/3}
%$$

\subsection{Democratization as brane accretion}

The CS-BH accretion effect  clarifies the puzzle of the micro-BH
non-democracy. The accretion process, as seen from the point of
view of the four-dimensional large distance BH physics, is the
process of democratization.  During this process, the BH gradually
loses the ability to distinguish among the different species,
i.e., looses its hair. As demonstrated above, this is a fully
classical process,  the end result of which is the static BH which
is pierced by all the strings. Such a BH interacts will all the
four-dimensional modes democratically. All of these modes share the
same horizon and correspondingly are emitted with the common
Hawking temperature.

Obviously,  democratization  is fully consistent with the
existence  of the {\it temporary}  hair,  that is suggested by the
unitarity arguments \cite{cutoff, GiaBHs}.  The democratization
time scale is automatically such, that unitarity is always
maintained in the BH production and evaporation processes.

In the above-studied codimension-2 system,  the democratization time is  the
same as the accretion time, which  for a given BH-CS system is
\begin{equation}
\label{demotime} t_d \, \sim \, z_0 \left( {z_0^3 \over  r_g^3
\delta} \right)^{{1\over 2}} \, ,
\end{equation}
where, $z_0$ is the initial BH-CS distance.\footnote{The extension
of this estimate to the branes of codimensions $q$ higher than 2 is
straightforward. The interaction potential in that case is
nontrivial already at the linear level, and is given by Eq. \eqref{VZ}.
 It follows from this potential, that the
democratization time scale is $ t_d\sim \lp(z_0^{q} / (T \, G_{4+n})\rp)^{1/2}$.
Note, that to this order, it is independent of the BH mass. }

On the other hand,  the evaporation
time of a six-dimensional BH localized on a given brane,  and
evaporating into  a single species is,
\begin{equation}
\label{evapor} t_{BH} \, \sim \, r_g^5/G_N^{(6)} \, .
\end{equation}
This time becomes comparable to the democratization (accretion)
time  $t_d$, when
\begin{equation}
\label{crossover} z_0 \,  = \, r_g(r_gM_*)^{{8\over 5}} \, ,
\end{equation}
where for definiteness,  we have taken $\delta\, \sim \, 1$. From
this equation, it  is  clear that the democratization process is
never in conflict with the non-universal nature of the quantum evaporation of the
small BHs, enforced by unitarity.  Indeed,  since for any
quasi-classical BH $(r_gM_*) \, \gtrsim 1$,  whenever  the
democratization time $t_d$ exceeds the evaporation time $t_{BH}$,
the BH and CS  are  separated by a distance much larger than the
BH gravitational radius,   $z_0  \, \gg  \, r_g$. Such a BH will
obviously evaporate before the democratization  process is
complete, and the species localized on the distant brane will not
be among the evaporation products. This is obviously consistent
with unitarity.  On the other hand,  the BHs for which $t_d \,
\lesssim \, t_{BH}$,  become democratic before they  have chance
to evaporate.  The decay process of such BHs will be universal in all the
localized species, and will not violate the univatiry  bound.

The above phenomenon demonstrates the  key point of our finding.
The existence of the classical democratization transition,
reconciles the non-\-democracy of the quasi-classical micro-BHs
(imposed by  the unitarity), with the universal thermal nature of
the Hawking radiation, according to which all the species should
see the same BH horizon, and be produced at the same thermal rate.
The effect of democratization  makes sure that BHs loose ability
to differentiate among the species within the finite time scale
$t_d$.   For the time scales $t \ll t_d$,   the system is {\it
classically}  time-dependent and BHs are free to enjoy the
non-democracy, without conflicting with the notion of universality
of the Hawking radiation.

Our resolution of the micro-BH non-democracy puzzle,  is not
limited to the extra dimensional theories, but should apply  to
general class of theories with multiple species.   The net result
then is, that in any theory with $N$ light species, the micro-BHs
have the two characteristic time scales. The first one, $t_{BH}$
is the usual quantum evaporation time, due to  Hawking radiation.
The second  time scale $t_d$ is a characteristic time of a purely
classical process of the BH democratization, during which the BH
looses its ability to differentiate between the different species,
and all the species start seeing the same BH horizon.

 \section{Implication for the Accelerator Experiments}

In any theory which includes Einsteinian  gravity as its low
energy limit, the ultimate consequence of the elementary particle
collisions, at sufficiently high center of mass energies and
sufficiently small impact parameter, is the production of BHs.
The threshold where the BH formation starts happening depends on
the parameter  $M_*$,  the scale where gravity is getting strong.
The interesting physical property of the large extra dimensional
framework is, that this scale can be reachable in the near-future
collider experiments. In such a case, the lightest  accessible
BHs will be essentially quantum \cite{AADD}. However,  production
of the larger quasi-classical BHs is also possible \cite{largebh,
giasergey}.

However,  as we know by now,  the primary  suppresser  of the
strong gravity scale,  is not a geometry, but the number of
species $N$. So  the similar properties should  be shared by the
BHs in all the theories with sufficiently large $N$,
irrespectively of their underlying geometric origin.

Thus, the general results of our paper are applicable not just to
extra dimensional theories, but to more general class with
particle species.  Properties of  the microscopic BHs that we have
uncovered, have direct observational consequences, and have to be
taken into the account in  their  experimental search. In
particular  our results indicate the following  picture.

First, in  agreement with the previous work,  the quasi-classical
BHs produced in the collision of elementary particles will start
out with the full memory of their `parent' species. However,  in
time they start loosing this memory {\it classically}  and become
democratic over a certain time scale. This classical
democratization process will compete with the quantum Hawking
evaporation and the outcome depends on the details of the
underlying theory. In any case,  a typical signature \cite{GiaBHs} is the
correlation between the softening of the amplitudes (characteristic to BH production (see \cite{soft}, and references therein) and the democracy of the evaporation products.

For example, in an extra dimensional scenario, in which other
species are localized on the nearby branes, the democratization
rate  can exceed the evaporation rate, and the latter process can
quickly become universal in all the species.  Of course, the whole
transition will proceed in an unitary way. Monitoring the level of
BH democracy and the evaporation rate can be an interesting
observational bridge between the Standard Model particles and the
hidden sector species.

\section*{Acknowledgements}

We thank  Jaume  Garriga,  Michele Redi and Sergey
Sibiryakov for discussions and comments.
This work is supported in
part  by David and Lucile  Packard Foundation Fellowship for
Science and Engineering, by NSF grant PHY-0245068 and by the
European Commission under the  ``MassTeV"  ERC Advanced Grant
226371.

\section*{Appendix}
\appendix
\addcontentsline{toc}{section}{Appendix}
\section{More on Large Extra Dimensions with $N$ Branes}

Let us describe in more detail  the models with large extra
dimensions and a number of branes in the bulk. For the moment, we
will keep full generality and assume that the bulk is $4+n$
dimensional. As mentioned before, we shall add $N$ 3-branes and we
populate them with the localized modes. Among these
modes there are the fluctuations of the brane positions.  These
already give $nN$ light species in the 4D effective theory.  In each brane we
have $n$ localized modes corresponding to
 Goldstone bosons of $n$ independent translations that are spontaneously broken by the brane.
  In the approximation of non-interacting branes, these would be exactly massless.
  Brane interactions are expected to generate masses to $Nn-n$ of them.   These masses will be below
  the cutoff, so all the modes will count in our argument.
 Thus, having only these modes is enough to illustrate our point.
The Hawking radiation into these modes translates into a thermal
excitation of the transverse displacement of the brane location,
which occurs when it crosses the BH horizon \cite{mining}.

The $N$ branes can be modelled as topological defects, made out of
some higher dimensional fields. For example, in the 6D case, we
could model them as Abelian-Higgs vortices, in which case one
needs a complex scalar and a vector, but these details are not
going to be important in our discussion.

In order to have these light modes interacting through gravity
only, we shall assume that branes are made out of independent
fields (one per brane) that couple to each other through gravity.  In this case, we need
of order $N$ independent $4+n$-dimensional fields.  For
simplicity, we can assume that the bulk theory consists of $N$
copies of the field theory that supports a brane as a smooth
solution. This immediately tells us that the actual gravitational
cutoff $\Lambda_*$ of the $4+n$ dimensional theory is
\cite{cutoff}
\begin{equation}\label{Lambda*}
  \Lambda_*^{2+n} \lesssim {M_*^{2+n} \over N}~.
\end{equation}

Note that the relation between the 4D Planck mass and the higher
dimensional one,
\begin{equation}\label{gauss}
M_P^2=R^n \; M_*^{2+n}
\end{equation}
%$$
%N_{KK} = (M_{*} R)^n~.
%$$
%the number of KK modes up to $M_*$.
is not changed by the presence of the branes, since it only relies
on the Gauss law.
However, in our setup, the actual number of KK modes (that is the
number of KK modes below the cutoff) is
\begin{equation}\label{NKK}
N_{KK} = ( \Lambda_* R)^n %= {N_{KK}\over N^{n\over n+2} }~.
\end{equation}
which is a factor $N^{-{n\over n+2}} $ smaller than without the
branes.

Note that Eq \eqref{gauss} can be written as
\begin{equation}\label{MPLambda}
M_P^2  \, \gtrsim \, N \, N_{KK} \;\, \Lambda_*^2~,
\end{equation}
which indicates that the total number of degrees of freedom is
$N_{total}= N\; N_{KK}$, as is indeed the case since each of the
$4+n$ dimensional fields gives rise to $N_{KK}$ KK modes. Note
that this counting of modes is quite independent of the actual
value of the masses of the 6D fields, as long as they are not too
close to $\Lambda_*$.

The main point concerning the Micro BH democratization transition
discussed above is unrelated to the hierarchy problem. However,
for completeness in this appendix we shall also consider the
possibility that the higher dimensional model solves the hierarchy
problem. This is accomplished by setting the cutoff
$\Lambda_*\simeq TeV$.
It is apparent from \eqref{MPLambda} that in this case, then the
number of modes in each KK tower is
\begin{equation}\label{NKKN}
N_{KK}\lesssim {10^{32} \over N}~.
\end{equation}

\subsection{Crowded extra dimensions}

In this Section, we describe the `crowded' limit of the ADD-type
models \cite{add} with $N$ branes, with the maximal possible
values of $N$. Let us start by obtaining the constraints arising
from including the branes as (solitonic) dynamical objects.
Generically, one can characterize the 3-branes by their tension,
$$
T \equiv \mu^4~,
$$
and their thickness, $\varepsilon$. Typically, in weakly coupled
field theories the tension scale $\mu$ is larger than
$1/\varepsilon$, but to simplify matters we will assume that they
are about the same, $\varepsilon \gtrsim 1/\mu$. Hence, one
obtains the thinnest possible branes for $\mu\lesssim \Lambda_*$.
So, the maximum number of branes that one can fit in the bulk
is\footnote{This assumes that the branes are not overlapping. In
principle one could relax this condition. However, we shall assume
it in order to have the different species localized at different
sites in the extra dimensions.}
\begin{equation}\label{Nmax}
N_{max} \simeq  (R \Lambda_*)^n = N_{KK}~.
\end{equation}
Hence, for a maximally populated bulk there are as many branes as
KK modes in the tower.\footnote{In principle, there are other
constraints on the total number of branes (or their tension)
coming from the gravitational effect of the branes. In the 6D
case, for example, this amounts to restricting the total tension
to be less than critical, $N \mu^4 < M_*^4$ (so that the total
deficit angle is less than $2\pi$). It is clear that this only
demands $\mu \lesssim \Lambda_*$, so at least in this case it does
not lead to any further constraint.}

As we already emphasized, requiring that the model solves the
hierarchy problem is orthogonal to the main point of this paper.
However, let us briefly consider this possibility.
First of all, the above arguments means that it suffices to have
of order $10^{16}$ high-dimensional species in order to solve the
hierarchy problem. Of course, this is completely equivalent to
having $10^{32}$ 4D species,  because each high dimensional field
brings in a tower of $10^{16}$ modes.
%Note that this is precisely the reason why the
%required number of high dimensional species is independent of the
%number of extra dimensions $n$.

Let us also point out that the phenomenological constraints on
models with $N$ bulk species are automatically milder as compared
to large extra dimensional models without them. The reason is that
the compactification radius must be smaller as compared to the
$N=1$ case. Combining \eqref{gauss} and \eqref{Lambda*}, one finds
\begin{equation}\label{R}
R^n = {M_P^2\over M_*^{n+2}} \leq {1\over N}{M_P^2\over
\Lambda_*^{n+2}}~.
\end{equation}
Setting $\Lambda_* \simeq TeV$ in order to solve the hierarchy
problem, this gives a compactification radius smaller than in the scenario
without the $N$ branes by a factor $N^{-n}$.

This is especially interesting for the $n=1$ case, since with the
maximal brane population this kind of (unwarped) model can be
rescued. Indeed, plugging in the numbers one finds that $R$ must
be millimetric or less in this case. Of course a more detailed
study of the phenomenology is required to see the viability of
this case.

\subsection{Micro Black Holes}

In general by `Micro' BHs we mean non-Schwarzschildian BHs, that
are quasi-classical solutions of the theory of gravity that operates at short distances $r < {\cal R}$.
Their masses are from around the cutoff $\Lambda_*$ to
some scale at least of order $\sqrt{N_{total}}~M_P$ \cite{GiaBHs}.
In the present theory there are two distinct kinds of Micro BHs.
The BHs with masses close enough to $\Lambda_*$ cannot behave as
semiclassical $4+n$ BHs, because in the high-dimensional theory
there is a large number of species as well and the cutoff correspondingly gets lowered
as compared to the
high-dimensional Planck mass.
However, for large enough masses of the  BHs, eventually a semiclassical high-dimensional
regime sets in.

\begin{figure}[t]
\vspace{-5mm}
\begin{center}
  \includegraphics[height=9cm]{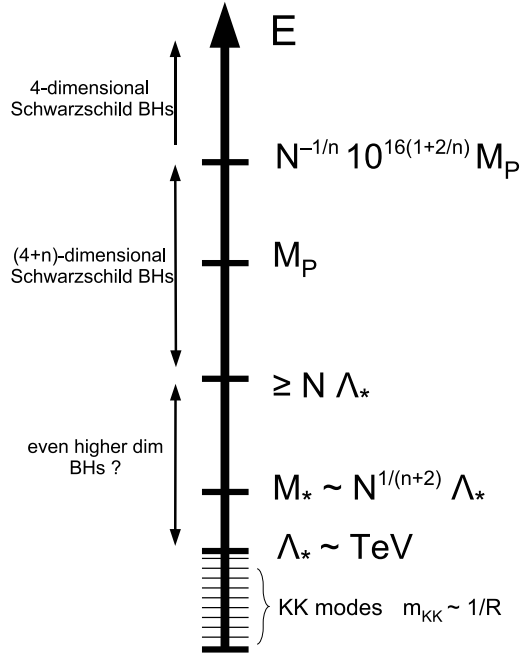}%
\end{center}
\caption{Schematic distribution of the relevant scales in the
large volume compactifications with $N$ branes. For small enough
masses, there are two types of non-Schwarzschildian BHs: for
moderate masses they are well approximated by $(4+n)-$dimensional
BHs. For smaller masses, the BHs are not even $(4+n)-$dimensional
Schwarzschildian. }
\label{fig:scales} %
\end{figure}

Arguing as in \cite{GiaBHs}, the threshold where the smallest
BHs will start behaving as quasi-classical high-dimensional BHs is
found by imposing that the $(4+n)-$dimensional relation between
the mass and radius, ${\cal M} = M_*^{n+2} \; {\cal R}_*^{n+1}$,
holds. Here, ${\cal R}_*$ denotes the BH radius for which the high
dimensional quasi-classical regime starts. The lowest possible
bound on ${\cal R}_*$ is $1/\Lambda_*$, so the smallest possible
window where the BHs are not even semiclassical $(4+n)-$ BHs is
$$
\Lambda_* \leq {\cal M}_{non-(4+n)\,Schwarzschild} \leq N
\Lambda_*~.
$$

On the other hand, the threshold of having the quasi-classical 4D BHs is
found by setting that the 4D mass-radius relation holds for the BH
horizon being equal to the compactification radius $R$. Hence the mass
of the smallest 4D BH is
\begin{equation}
M_P^2 \, R \, \lesssim \; N^{-{1\over n}} \lp({M_P\over\Lambda_*}\rp)^{n+2\over n} M_P %
\;\simeq \; N^{-{1\over n}} \;10^{16\,(1+{2\over n})} \;M_P
\end{equation}
%\begin{equation}
%M_P^2 \, R \lesssim M_P^2 \; {M_P\over N^{1/2} \Lambda_*^2}
%\;\sim\; N^{1/2} \,N_{KK} \; M_P
%\end{equation}
where we used \eqref{R} and \eqref{Lambda*}, and in the last
equation we assumed that $\Lambda_* \simeq TeV$. Note that (as
it happens in the large extra dimensional model without branes) this is a factor
$N_{KK}^{1/n}$ above the absolute lower bound for this scale,
$N_{total}^{1/2} M_P$
\cite{GiaBHs}.\\

%Hence, the window where we have approximately 6D-Schwarzschild BHs
%is
%$$
%N \Lambda_* \leq {\cal M}_{6D\,Schwarzschild} \leq N^{1/2}
%\,N_{KK} \; M_P~.
%$$
Hence, the window where we have an approximately
$(4+n)$-Schwarzschild BHs is
$$
N \Lambda_* \leq {\cal M}_{(4+n)\,Schwarzschild} \leq \;
%N^{-{1\over n}} \;10^{16\,(1+{2\over n})} \;M_P~.
\;10^{16} \lp( {10^{32} \over N} \rp)^{1/n}  \;M_P~.
$$
The resulting set of scales is arranged as shown in
Fig~\ref{fig:scales}.
Notice that in the limit $N = 1$, we recover the usual large extra
dimensional picture, where the high-dimensional BH window  is $M_*
\leq {\cal M}_{(4+n)\,Schw} \leq \,N_{KK}^{{1\over n}+{1\over2}}
\; M_P $.
Notice as well, that for the maximal number of branes allowed,
$N\sim 10^{16}$, the mass of the heaviest
non-($4+n$)-Schwarzschildian BH reaches out to precisely $M_P$.
However, there is  always a higher dimensional window.

\section{Brane-BH effective potential}
\label{sec:braneBHpot}

In this Appendix we derive the force between a BH and a test
brane, and show that it takes the form advanced in Section
\ref{sec:BHbackground}. For the sake of generality, we shall
consider Nambu-Goto branes of arbitrary dimension in a generic
$(4+n)$-dimensional bulk spacetime. This discussion follows
\cite{escape2,escape1}.

Let us start with a test 3-brane in the background given by a
$(4+n)$-dimensional Schwarzschild BH. To better visualize the
brane embedding in the BH background it is convenient to use the
so-called isotropic coordinates,
$$
ds^2_{4+n}=-f^2(R)dt^2+h^2(R)[d{\bf x}^2+d{\vec{z}}^{~2}]
$$
where ${\bf x}$ and ${\vec{z}}$ are `cartesian' 3-dimensional  and
$n-$dimensional coordinates respectively,
$$
R^2= r^2+z^2~, \qquad r^2\equiv {\bf x}^2~,\qquad z^2\equiv
{\vec{z}}^{\,2}
$$
and
$$
f(R)={1-{1\over4}\big({r_g\over R}\big)^{n+1} \over
1+{1\over4}\big({r_g\over R}\big)^{n+1} } \qquad
h(R)=\Big(1+{1\over4}\Big({r_g\over R}\Big)^{n+1} \Big)^{2\over
n+1}
$$
with $r_g$ the usual Schwarzschild radius,
$$
r_g^{n+1}={16\pi G_{4+n}\over (n+2)\Omega_{n+2}}M_{BH}
$$
and $\Omega_{n+2}$ the volume of an unit $(n+2)$-dimensional
sphere.

The brane embedding in this space can be parameterized as
$\vec{z}=\vec{z}(x^\mu)$, with $x^\mu=\{t,{\bf x}\}$ the
coordinates along the brane.
%
%
%$$
%\vec{z}=\vec{z}(x^\mu)~, \quad {\rm with} \quad x^\mu=\{t,{\bf
%x}\}~.
%$$
This leads to the following induced 4D geometry
$$
ds^2_{4}=ds_{(0)}^2 + h^2 (\partial_\mu\vec{z}\;dx^\mu)^2~,%
\qquad {\rm with} \qquad %
ds_{(0)}^2\equiv -f^2 dt^2+ h^2 d{\bf x}^2
$$
and here $f$ and $h$ evaluated at $R=\sqrt{r^2+z^2(x^\mu)}$.

In the probe approximation, the brane dynamics is described by the
Nambu-Goto action,
$$
S^{NG}=-T \int d^{4}x\; \sqrt{-g_4}=-T \int d^4x\;
f\,h^3\;\sqrt{1-h^2 g_{(0)}^{\mu\nu}
\partial_\mu \vec{z} \cdot \partial_\nu \vec{z} }~,
$$
where as usual $T$ is the tension.
The generalization to a $p+1$ dimensional brane is
straightforward,
\begin{equation}\label{SNG}
S^{NG}_p=-T \int d^{p+1}x\; f\,h^p\;\sqrt{1-h^2 g_{(0)}^{\mu\nu}
\partial_\mu \vec{z} \cdot \partial_\nu \vec{z} }
\end{equation}
where the number of $\vec z$ coordinates equals the
\emph{codimension},
$$
q\equiv n+3-p~.
$$

The energy of static configurations is
\begin{equation}\label{Epq}
E[\,\vec{z}(x)\,]=T \int
d^px\,\lp(1-\psi \rp) \lp(1+\psi\rp)^{{p+2-q\over p-2+q}}%
\;\sqrt{1+\partial^i \vec{z} \cdot \partial_i \vec{z} }
\end{equation}
where we introduced the notation
$$
\psi\equiv{1\over4}{r_g^{n+1}\over (r^2+z^2)^{n+1\over2}}~.
$$
%\begin{equation}\label{Ep2}
%E_{p,2}(z)=T\int d^px\,\lp(1-\psi^2 \rp) \qquad\qquad v=-\psi^2~,
%\end{equation}
%which as expected vanishes at linear order in $G_N$ (this being
%second order only is a peculiarity of the isotropic coordinates).

Far enough from the BH, the gravitational potential is small,
$\psi \ll1$, everywhere on the brane and the gradients of $\vec z$
are also small. Up to a (divergent, but irrelevant) constant, the
energy of the brane configurations is
\begin{equation}\label{Esmall}
E[\,\vec{z}(x)\,]\simeq T \int d^px\,%
\lp(  -2{q-2\over p+q-2} \psi - {1\over2}
\vec{z}\cdot\Delta\vec{z} \rp) + {\rm const}
\end{equation}
where higher order terms in the (high-dimensional) Newton constant
$G_{4+n}$ are neglected. For codimension 2, the potential energy
term in \eqref{Esmall} vanishes at linear order $G_{4+n}$. This is
of course a manifestation of the fact  that codimension-2 branes do not  either attract
or repel the test particles. In this case,  we have to include in
the expansion of the energy the next order  terms. Thus, for codimension
2 we have
\begin{equation}\label{Ecod2}
E_{cod2}[\,\vec{z}(x)\,]\simeq T \int d^px\,%
\lp(  - \psi^2 - {1\over2} \vec{z}\cdot\Delta\vec{z} \rp) + {\rm
const}
\end{equation}

Recall that we are interested in the potential energy of the brane
configurations that are asymptotically separated from the
`equatorial' plane of the BH by a certain distance $Z$. As done in
\cite{escape2}, one way to obtain this is  the following: we first
solve for the profile of the bran with the boundary condition that
$|\vec{z}|$ approaches a constant ($Z$) at infinity ($r\to\infty$)
and then evaluate the energy of these configurations.

The first step requires to extremize the functional \eqref{Ecod2}.
Denoting by $z$ the non-zero component of $\vec{z}$, the
extremization leads to $\Delta z = v_{,z}$ with $v(r,z)$ the
potential appearing in \eqref{Esmall} and \eqref{Ecod2}. Hence,
the energy of the minimum energy configurations is of the form $ T
\int d^px \,[v-(1/2)z\,v_{,z}] $ with $v$ and $v_{,z}$ evaluated
on the profile $z(r)$. However, if we are interested in the
potential to leading order in $G_{4+n}$, it suffices to replace
the actual profile by the constant $Z$ to which $z(r)$ asymptotes.
Thus, for $q\neq2$ the leading order potential is
\begin{equation}\label{EZsmall}
E(Z)\simeq - \lp(
{q\over8}\,{\Gamma(p/2)\Gamma(q/2)\over\Gamma[(p+q)/2]
}\;\Omega_{p-1} \rp)\; {T \, r_g^{n+1} \over Z^{q-2}}
\end{equation}
Eq \eqref{EZsmall} encodes the well known gravitational effect of
branes of different codimension: for codimension 3 or more, the is
an attractive force, for codimension 1 it is repulsive, whereas
for codimension 2 there is no force.

Hence, what sets the sign of the force for $q=2$ is the next term,
of order $G_N^2$. One obtains
\begin{equation}\label{EZcod2}
E_{cod2}(Z)\simeq %
-\lp({\Gamma(p/2)\Gamma(2+p/2)\over 16\Gamma(p+1) }%
\;\Omega_{p-1}\rp) %
{T \, r_g^{2(n+1)} \over Z^{p}} \, ,
\end{equation}
which gives an attraction. These results reproduce the behaviour
quoted in Section \ref{sec:BHbackground}.

\end{document}